\documentclass[12pt]{amsart}
\usepackage{color} 
\usepackage{amsmath}
\usepackage{amsxtra}
\usepackage{amscd}
\usepackage{amsthm}
\usepackage{amsfonts}
\usepackage{amssymb}
\usepackage{eucal}
\usepackage{epsfig}
\usepackage{graphics}
\def\({\left(}
\def\){\right)}

\newcommand{\nn}{\nonumber}
\newcommand{\bea}{\begin{eqnarray}}
\newcommand{\ena}{\end{eqnarray}}
\def\bel{\begin{eqnarray}}
\def\enl{\end{eqnarray}}
\newcommand{\be}{\begin{eqnarray*}}
\newcommand{\en}{\end{eqnarray*}}
\newcommand{\ba}{\begin{array}}
\newcommand{\ea}{\end{array}}

\newcommand{\slt}{\mathfrak{sl}_2}

\newenvironment{tenumerate}{
  \begin{enumerate}
  
  }{\end{enumerate}}
\newcommand{\bi}{\begin{tenumerate}}
\newcommand{\ei}{\end{tenumerate}}
\newcommand{\isoto}[1][]%
{{\mathop{\buildrel{\sim}\over\longrightarrow}\limits_{#1}}}

\def\[{\left[}
\def\]{\right]}

\newcommand{\al}{\alpha}

\numberwithin{equation}{section}

\def\Ja{J_{\mathrm{aff}}}

\def\bi{\mathbf{i}}

\definecolor{5/18}{rgb}{0.9,0,0.7}
\definecolor{5/19}{rgb}{0.9,0,0.7}
\definecolor{5/19comment}{rgb}{0.9,0.7,0}
\definecolor{ao}{rgb}{0.0, 0.5, 0.0}

\newcommand*{\relphantom}[1]{\mathrel{\phantom{#1}}}
\DeclareMathOperator{\ch}{ch}
\DeclareMathOperator{\res}{res}

\begin{document}

\begin{title}[Problems with using separated variables]
{Problems with using separated variables for computing expectation values for higher ranks}
\end{title}
\author{D.~Martin and F.~Smirnov}
\address{DM\footnote{The work was conducted while this author was doing an internship at LPTHE}:
D{\'e}partment de Physique, {\'E}cole Normale Sup{\'e}rieure,
75005 Paris, France}\email{david.martin2@ens.fr}
\address{FS\footnote
{Membre du CNRS}: 
Laboratoire de Physique Th{\'e}orique et
Hautes Energies, Universit{\'e} Pierre et Marie Curie,
Tour 13, 4$^{\rm er}$ {\'e}tage, 4 Place Jussieu
75252 Paris Cedex 05, France}\email{smirnov@lpthe.jussieu.fr}

\begin{abstract}
We consider the simplest classical integrable model corresponding to a
non-hyperelliptic spectral curve. We show that a certain complicated integral
occurs when computing the average of observables in this model. This integral
does not factorise. Since similar problems should also exist in the quantum
case, we think that a serious question arises of how to deal with these
integrals.
\end{abstract}

\maketitle

\section{Introduction}

The relation of integrable models to the affine Jacobi variety is explained by
Mumford~\cite{mum} for the hyperelliptic case. The general framework is as
follows: we start with a matrix $l(z)$ depending on a spectral parameter $z$. A certain Poisson bracket is defined such that the coefficients
of the characteristic polynomial $R(z,w)=\det\big(wI+l(z)\big)$ are in involution. The
Hamiltonian of the system is one of those. Then the ring of functions generated
by the coefficients of $l(z)$ is the affine ring of the affine Jacobi variety
$\Ja$ of the algebraic curve $\Sigma$ defined by $R(z,w)=0$. Let $g$ be the
genus of this curve, and denote the points on the curve by $P=(z,w)$.

The coefficients  of $l(z)$ can be expressed in terms of a divisor
$(P_1,\cdots,P_g)$, the corresponding $z_i$, $w_i$ being canonical with respect
to the Poisson structure. These are the separated variables of Flaschka and
McLaughlin~\cite{FM}. The integrable model comes with some reality condition. In
the language of algebraic geometry, this means that some half-basis $a_1,\cdots,a _g$ of
homologies  is chosen. The angle variables $\varphi_j$ are
obtained by Abel transformation, which uses the holomorphic differentials
normalised with respect to the chosen half-basis. The Abel transformation
requires choosing an initial point on the surface. We shall assume for
simplicity that there is only one infinite point $\infty$ on the curve, and we
take it as the initial point.

There is a problem which does not look like the most important one for the
classical case, but becomes crucial after quantisation. Consider the average of
any observable $x$ (element of the affine ring algebraic-geometrically) over a
classical trajectory for a long period of time. By ergodic considerations, in
generic position this is the same as the average over the torus of angle
variables:
\begin{equation*}
  \langle x\rangle
  =\frac 1 {(2\pi)^g}\int\limits_0^{2\pi}\cdots\int\limits_0^{2\pi}
   x(\varphi_1,\cdots,\varphi _g)\,d\varphi_1\wedge\cdots\wedge d\varphi_g\,.
\end{equation*}
This integral looks rather complicated having in mind that the typical
expression for $x$ is
\begin{equation*}
  x(\varphi)=\frac{\Theta_k(\varphi)}{\theta\big(\varphi+\Delta(\infty)\big)^k}\,,
\end{equation*}
where $\theta(\varphi)$ is the Riemann theta function, $\Delta(\infty)$ is the
Riemann characteristic corresponding to our choice of the initial point and
$\Theta_k(\varphi)$ is an arbitrary theta function of order $k$.

In order to simplify the problem we undo the Abel transformation
following~\cite{SmToda}:
\begin{equation}
  \langle x\rangle
  =\frac 1{\mathcal{N}}\int\limits_{a_1}\cdots\int\limits_{a_g}x(P_1,\cdots,P_g)
   \,d\sigma_1\wedge\cdots\wedge d\sigma _g\,.
\label{aver}
\end{equation}
We use the same notation for the observable $x$, but now it depends on the
divisor. Set
\begin{equation*}
  d\sigma_j=\sum_{k=1}^gd\sigma_j(P_k)\,,
\end{equation*}
where the $d\sigma_j(P)$ are holomorphic differentials. They are not necessarily
normalised, so the denominator
\begin{equation*}
  \mathcal{N}=\det\(\int\limits_{a_j}d\sigma_k(P)\)_{j,k=1,\cdots,g}
\end{equation*}
was included.

We identify the functions on $\Ja$ with top forms via
\begin{equation*}
  x(P_1,\cdots,P_g)\ \longrightarrow \
  x(P_1,\cdots,P_g)\,d\sigma_1\wedge\cdots\wedge d\sigma_g\,,
\end{equation*}
and the affine ring is generated by the functions
\begin{equation}
  \frac{d\xi_1\wedge\cdots\wedge d\xi_g}{d\sigma_1\wedge\cdots\wedge d\sigma_g}\,.
  \label{lin}
\end{equation}
Here the $d\xi_j$ are arbitrary differentials with singularity at $\infty$. So
$x$ may contain in the denominator the determinant of
$d\sigma_1\wedge\cdots\wedge d\sigma_g$ in any power. This is what makes the
integral complicated.

Notice however that these are not functions, but rather top cohomologies which
we are interested in when computing the integral~\eqref{aver}. The simplest case
occurs in the study of the Toda chain, KdV equation, sine-Gordon equation
\emph{etc}. There the underlying finite-dimensional algebra is $\slt$, and the
spectral curve is hyperelliptic. For the hyperelliptic case it was
shown~\cite{NS,nak} that the top cohomologies can be realised as functions which are
linear in~\eqref{lin}. Since the integrand in ~\eqref{aver} can be reduced to 
a linear combination of cohomologies for any $x(P_1,\cdots,P_g)$, we conclude that
$\langle x\rangle$ is always given by a sum of ratios of determinants of
one-fold integrals (the denominator is always $\mathcal{N}$). Furthermore, it is
possible to describe the averages in a more universal way using the canonical
second kind differential $\omega(P_1,P_2)$ defined on $\Sigma\times\Sigma$ with
vanishing $a$-periods and a quadratic singularity at $P_1=P_2$.

In the papers~\cite{SZ,NS1}, the cohomologies of the affine Jacobi variety were
studied for spectral curves corresponding to algebras of higher rank. In this
case the spectral curves are $N$-fold coverings of the Riemann sphere, and it
was shown that the structure of cohomologies is much more complicated than in
the hyperelliptic case. The goal of the present paper is to explain this point
on the simplest possible non-hyperelliptic example: three-sheet covering of the
sphere of genus $3$. The reason for considering this case is that the
papers~\cite{SZ,NS1} are rather complicated to follow, and they lack a clear
example of an observable $x$ for which the integral~\eqref{aver} does not
factorise. Let us be more explicit: The spectral curve is given by
\begin{equation*}
  w^3+w^2t_1(z)+wt_2(z)+z^4+t_3(z)=0
\end{equation*}
with $\deg_z{\big(t_j(z)\big)}=j$. Define
\begin{equation}
  \(z^{k_1}w^{l_1},z^{k_2}w^{l_2},z^{k_3}w^{l_3}\)
  =\left|\begin{matrix}
     z_1^{k_1}w_1^{l_1}&z_2^{k_1}w_2^{l_1}&z_3^{k_1}w_3^{l_1}\\
     z_1^{k_2}w_1^{l_2}&z_2^{k_2}w_2^{l_2}&z_3^{k_2}w_3^{l_2}\\
     z_1^{k_3}w_1^{l_3}&z_2^{k_3}w_2^{l_3}&z_3^{k_3}w_3^{l_3}
   \end{matrix}\right|\,.
  \label{def()}
\end{equation}
Then, starting from observables $x$ quadratic in coefficients of $l(z)$, the
integral
\begin{equation}
  I=\int\limits_{a_1}\int\limits_{a_2}\int\limits_{a_3}\frac{(1,z,w^2)^2}{(1,z,w)}
    \prod_{j=1}^3\frac{dz_j}{\partial_{w_j}R(z_j,w_j)}
  \label{integral}
\end{equation}
occurs, which cannot be reduced to a factorised form by adding total derivatives
to the integrand.

In this paper we consider linear Poisson brackets, while in many applications
quadratic ones are necessary. This does not change much in the
algebraic-geometrical approach, though.

Let us say a few words about quantisation. The most convenient framework for
investigating correlation functions of quantum integrable models is the
so-called ``quantum transfer-matrix method'' in~\cite{Suzuki,Klumper}. In the
paper~\cite{HGSIII} it is called the Matsubara method, because in the
relativistic case it is clearly equivalent to the Matsubara trick. The
most important tool of this method is the computation of vacuum expectation values for
elements of $\mathbf{A}_\mathrm{qua}$, the algebra of all products of elements
of the monodromy matrix (analogue of $l(z)$) in Matsubara direction. Then one
can try to understand the algebraic structure behind the expressions for these
expectation values. In the case whose classical limit corresponds to
hyperelliptic spectral curves, the  algebra $\mathbf{A}_\mathrm{qua}$ allows
fermionic structure~\cite{HGS,HGSII}, and the expectation values are expressed
in terms of a quantum deformation of the canonical second kind differential
$\omega(P_1,P_2)$. This is explained by the fact that in the quantum case one
can use the method of separation of variables~\cite{Skl,SmToda} for computing
the expectation values, which provides the direct analogue of~\eqref{aver}.
The relation to the separation of variables is not explicit
in~\cite{HGSIII}, but it was used there as a heuristic argument.

Vaguely, the structure of matrix elements in our case of three degrees of
freedom (genus) is
\begin{equation}
  \langle x\rangle
  =\int\limits_{C_1}dz_1\int\limits_{C_2}dz_2\int\limits_{C_3}dz_3\prod_{i=1}^3
   Q(z_i)x(z_1,w_1,z_2,w_2,z_3,w_3)\prod_{j=1}^3Q(z_j)\varphi(z_j)\,,
\end{equation}
where the $C_j$ are some contours in the quantum plane, $Q(z)$ is the vacuum
eigenvalue of the Baxter $Q$-operator and $\varphi(z_{j})$ is a certain function
similar to the one used in~\cite{HGSIII}. The classical limit of
$\prod_{j=1}^3Q^2(z_j)\varphi(z_{j})$ coincides with
$\prod_{j=1}^3\frac{dz_j}{\partial_{w_j}R(z_j,w_j)}$. The denominator of
$x(z_1,w_1,z_2,w_2,z_3,w_3)$, which was already hard in the classical case,
becomes very strange after quantisation having in mind that in
$z$-representation the $w_j$ are shifts or derivations depending on the
particular model. Wishing to apply separation of variables to compute matrix
elements for higher ranks, one has to learn how to work with integrals for such
$x(z_1,w_1,z_2,w_2,z_3,w_3)$. This is the main conclusion of the present paper.

\section{The model}

Consider the Lie algebra $\mathfrak{g}=\mathfrak{sl}_3$ and the corresponding
loop algebra in principle gradation $\hat{\mathfrak{g}}_0$. The coadjoint
representation $\hat{\mathfrak{g}}_0^*$ allows the Kirillov-Kostant Poisson
structure which we would like to describe using the $r$-matrix formalism.
Namely, consider a $3\times 3$ matrix $m(z)$ polynomially depending on the
spectral parameter $z$. We write down its Poisson brackets using usual tensor
notations as
\begin{equation}
  \{m_1(z_1),m_2(z_2)\}=[r_{1,2}(z_1,z_2)\ ,\ m_1(z_1)+m_2(z_2)]\,,
  \label{PB0}
\end{equation}
where
\begin{align*}
  &r_{1,2}(z_1,z_2)=\frac 1 {z_1-z_2}\(\frac 1 2(z_1+z_2)t^{00}+z_1t^{-+}
                    +z_2t^{+-}\)\,,\\
  t^{00}=\sum&E_{i,i}\otimes E_{i,i}\,,\qquad
  t^{{-+}}=\sum_{i>j} E_{i,j}\otimes E_{j,i}\,,\qquad
  t^{{+-}}=\sum_{i<j} E_{i,j}\otimes E_{j,i}\,,
\end{align*}
with $E_{i,j}$ being the $3\times 3$ matrices with 1 at position $(i,j)$ and 0
everywhere else.

The finite-dimensional polynomial orbits are of great interest. Their structure
must be compatible with the bracket above. In this paper we shall consider the
simplest non-trivial case corresponding to the spectral curve of genus $g=3$:
\begin{equation*}
  m(z)
  =\begin{pmatrix}
     m_{1,1}^{(0)}z+m_{1,1}^{(1)}& m_{1,2}^{(0)}z+m_{1,2}^{(1)}& m_{1,3}^{(1)}\\
     m_{2,1}^{(1)}z& m_{2,2}^{(0)}z+m_{2,2}^{(1)}& m_{2,3}^{(0)}z+m_{2,3}^{(1)}\\
     m_{3,1}^{(0)}z^2+m_{3,1}^{(1)}z& m_{3,2}^{(1)}z&m_{3,3}^{(0)}z+ m_{3,3}^{(1)}
\end{pmatrix}\,.
\end{equation*}
There is one simpler case with $g=1$, but being elliptic it is of no interest to
us; it has been said that we want to consider non-hyperelliptic curves only. The
element $m_{1,2}^{(0)}m_{2,3}^{(0)}m_{3,1}^{(0)}$ lies in the centre of the
Poisson algebra, so we will set it equal to $1$.

\section{Spectral curve}

Consider the characteristic polynomial
$$R(z,w)=\det\big(wI+m(z)\big)\,.$$
The crucial role in the investigation of the model is played by the spectral
curve
\begin{equation}
  R(z,w)=0\,,
  \label{curve1}
\end{equation}
which in our case reads explicitly as
\begin{equation}
\begin{split}
  R(z,w)
  &=w^3+w^2\(t_1^{(1)}z+t_1^{(2)}\)+w\(t_2^{(1)}z^2+t_2^{(2)}z+t_2^{(3)}\)\\
  &\relphantom{=}+z^4+t_3^{(1)}z^3+t_3^{(2)}z^2+t_3^{(3)}z+t_3^{(4)}\,.
  \label{curve2}
\end{split}
\end{equation}

Before going any further, let us introduce a basis of Abelian differentials on
the curve. The differentials, singular at $\infty$ (this point) only, are of the
form
\begin{equation*}
  d\sigma(P)=\frac{Q(z,w)}{\partial_{w}R(z,w)}\,dz\,,
\end{equation*}
where $Q(z,w)$ is a polynomial. For differentials with no residue at $\infty$
(which is definitely the case for differentials with singularity at this point
only), the primitive function is well-defined in the vicinity of $\infty$. So
there is a pairing
\begin{equation*}
  d\omega_1\circ d\omega_2=\res_{z=\infty}\bigl(d\omega_1\omega_2\bigr)\,,
\end{equation*}
and we want to define the canonical basis with respect to it. To this end
introduce the intersection form
\begin{equation*}
  C(P_1,P_2)
  =d\!\(\frac{R(z_1,w_2)}{(z_1-z_2)(w_1-w_2)}\frac{dz_2}{\partial_{w_2}R(z_2,w_2)}
   \!+\!\frac{R(z_2,w_1)}{(z_2-z_1)(w_2-w_1)}\frac{dz_1}{\partial_{w_1}R(z_1,w_1)}\)
\end{equation*}
satisfying
\begin{equation*}
  \int\limits_{\gamma_1}\int\limits_{\gamma_2}C(P_1,P_2)
  =2\pi i\,\gamma_1\circ\gamma_2
\end{equation*}
for any two cycles $\gamma_1$, $\gamma_2$. We have
\begin{equation*}
  C(P_1,P_2)=\sum_{j=1}^3\big(d\sigma_j(P_1)\,d\tilde{\sigma}_j(P_2)
             -d\sigma_j(P_2)\,d\tilde{\sigma}_j(P_1)\big)\,,
\end{equation*}
where the polynomials $Q(z,w)$ for holomorphic differentials $d\sigma_j$  and
their dual second kind differentials $d\tilde{\sigma}_j$ are as follows:
\begin{align}\label{Q}
  &Q_1(z,w)=1,\quad
  &\widetilde{Q}_1(z,w)
   &=5wz^2+3z^3t^{(1)}_1-w^2t^{(1)}_2+wz\(3t^{(1)}_3-t^{(1)}_1t^{(1)}_2\)\\
  &&&\relphantom{=}+z^2\(2t^{(1)}_1t^{(1)}_3+2t^{(2)}_1-\(t^{(1)}_2\)^2\)\,,\nn\\
  &Q_2(z,w)=z,
  &\widetilde{Q}_2(z,w)
   &=2wz+z^2t^{(1)}_1-t^{(1)}_3t^{(2)}_1+t^{(1)}_2t^{(2)}_2-t^{(1)}_1t^{(2)}_3\,,\nn\\
  &Q_3(z,w)=w,
  &\widetilde{Q}_3(z,w)
   &=z^2+t^{(1)}_2t^{(2)}_1-t^{(2)}_3\,.\nn
\end{align}
The pairings are  canonical, i.e.\
\begin{equation*}
  d\sigma_i\circ d\sigma_j=d\tilde\sigma_i\circ d\tilde\sigma_j=0,\qquad
  d\sigma_i\circ d\tilde\sigma_j=\delta _{i,j}\,,
\end{equation*}
or in other words, $C(P_1,P_2)$ is invariant under the action of the modular group.

\section{Reduction and affine ring}\label{reduction}

The coefficients of the characteristic polynomials are in
involution:
\begin{equation*}
  \{R(z_1,w_1),R(z_2,w_2)\}=0\,.
\end{equation*}
Setting them to constants we obtain a $5$-dimensional variety. This is not quite
satisfactory, because we want this level variety to be equivalent to the affine
Jacobi variety, i.e.\ to have dimension $3$. Thus there are two degrees
of freedom which we want to eliminate. This is basically the same procedure as
in~\cite{SZ}, but there are some differences due to a different structure of the
principal part of $l(z)$.

We have
\begin{equation*}
  m(z)=z^2m_{3,1}^{(0)}E_{{3,1}}+z\mu +O(1)\,.
\end{equation*}
Let $\mu _1$ be the first line of $\mu$ and introduce the matrix
\begin{equation*}
  s=\begin{pmatrix}1&0&0\\ &\mu_1\\ &\mu_1\mu\end{pmatrix}\,.
\end{equation*}
Then it is easy to compute that the modified matrix
\begin{equation*}
  l(z)=sm(z)s^{-1}
\end{equation*}
satisfies the Poisson brackets
\begin{equation}
  \{l_1(z_1),l_2(z_2)\}
  =[\tilde r_{1,2}(z_1,z_2),l_1(z_1)]-[\tilde r_{2,1}(z_2,z_1),l_2(z_2)]
  \label{modPB}
\end{equation}
with
\begin{equation*}
  \tilde r_{1,2}(z_1,z_2)=\frac {z_2}{z_1-z_2}P_{1,2}+z_2u_{1,2}\,,
\end{equation*}
where $P_{1,2}=t^{00}+t^{+-}+t^{-+}$ is the permutation and
\begin{equation*}
  u_{1,2}
  =E_{2,1}\otimes E_{3,1}+E_{3,1}\otimes E_{3,2}+E_{3,2}\otimes E_{3,1}\,.
\end{equation*}
Hence the new $L$-operator has the form
\begin{equation}
  l(z)=\begin{pmatrix}
         l^{(1)}_{1,1}&z+l^{(1)}_{1,2}&l^{(1)}_{1,3}\\
         l^{(1)}_{2,1}&l^{(1)}_{2,2}&z+l^{(1)}_{2,3}\\
         z^2+zl^{(0)}_{3,1}+l^{(1)}_{3,1}&zl^{(0)}_{3,2}+l^{(1)}_{3,2}
         &zl^{(0)}_{3,3}+l^{(1)}_{3,3}
       \end{pmatrix}\,.
  \label{Lop}
\end{equation}
Now the level of $R(z,w)$ is three-dimensional and it coincides with the affine
Jacobi variety
\begin{equation*}
  \Ja=J-\Theta\,,
\end{equation*}
where $J$ is the Jacobi variety of the curve and $\Theta$ is the theta divisor.

We realise the Jacobi variety via divisors $(P_1,P_2,P_3)$.
For example, by using notation~\eqref{def()}, the wedge product of the first
kind differentials equals
\begin{equation*}
  (1,z,w)\cdot\frac{{d z_1}\wedge dz_2\wedge dz_3}{\prod_j\partial_{w_j}R(z_j,w_j)}\,.
\end{equation*}
The ring of functions on $\Ja$ (affine ring) is generated by
\begin{equation}
  \frac{\(z^{k_1}w^{l_1},z^{k_2}w^{l_2},z^{k_3}w^{l_3}\)}{(1,z,w)}\,,
  \label{form}
\end{equation}
and we shall see that it is actually sufficient to take the following six of
them:
\begin{equation*}
  \frac{\(1,z,z^2\)}{(1,z,w)},\quad \frac{\(1,w,z^2\)}{(1,z,w)},\quad
  \frac{(1,z,zw)}{(1,z,w)},\quad \frac{(1,w,zw)}{(1,z,w)},\quad
  \frac{\(1,z,w^2\)}{(1,z,w)},\quad \frac{\(1,w,w^2\)}{(1,z,w)}\,.
\end{equation*}

\vskip .2cm
\noindent\textbf{Remark.} \emph{Notice that the filtration by the degree of
$(1,z,w)$ in the denominator is not exactly the same as the filtration by the
order of the pole on the theta divisor, because the latter contains also the
points $(P_1,P_2, P_3)$ with $P_k=\infty$. For instance, the function
$z_1+z_2+z_3$ has a pole on the theta divisor, but no denominator.}
\vskip .2cm

On the other hand, the affine ring $\mathbf{A}$ coincides with the polynomial
ring of all coefficients of $l(z)$~\eqref{Lop}. The most important feature of
this analysis consists in the possibility to introduce a grading on that ring.
This grading is very different from general algebraic-geometrical means. It is
introduced by the two conditions
\begin{equation*}
  \deg{(z)}=3\,, \qquad \deg{\(l(z)_{i,j}\)}=4+i-j\,.
\end{equation*}
For instance, $\deg{\(l^{(1)}_{1,1}\)}=4$ and $\deg{\(l^{(1)}_{3,2}\)}=5$.
Hence the consistent degree of $w$ is $\deg{(w)}=4$. The grading implies the decomposition
\begin{equation*}
  \mathbf{A}=\bigoplus_{p=0}^{\infty}\mathbf{A}^{(p)}\,,
\end{equation*}
where $\mathbf{A}^{(p)}$ are the subspaces of degree $p$. We find for the
character:
\begin{equation*}
  \ch{(\mathbf{A})}
  =\sum_{p=0}^\infty q^p\dim{\(\mathbf{A}^{(p)}\)}
  =\frac 1 {(1-q)(1-q^2)^2(1-q^3)^3(1-q^4)^3(1-q^5)^2(1-q^6)}\,.
\end{equation*}

Consider now the ring $\mathbf{F}$ generated by the coefficients of $R(z,w)$.
Its character is
\begin{equation*}
  \ch{(\mathbf{F})}
  =\frac 1 {(1-q)(1-q^2)(1-q^3)(1-q^4)(1-q^5)(1-q^6)(1-q^8)(1-q^9)(1-q^{12})}\,.
\end{equation*}
Finally introduce the ring
$$\mathbf{A}_0=\mathbf{A}/\mathbf{F}\mathbf{A}$$
with character
\begin{equation}
  \ch{(\mathbf{A}_0)}
  =\frac{\ch{(\mathbf{A})}}{\ch{(\mathbf{F})}}
  =\frac{(1+q^4)(1+q^3+q^6)(1+q^4+q^8)}{(1-q^2)(1-q^3)(1-q^5)}\,.
  \label{chA0}
\end{equation}

The last formula means that the map
$m\colon\mathbf{A}_0\otimes\mathbf{F}\to\mathbf{A}$ has no kernel. Like
in~\cite{NS1}, we give two arguments for that. The informal argument is that
the existence of a kernel would contradict the fact that over $\mathbb{C}$, the
level of $t^{(p)}_k$ is generically a $g$-dimensional variety. The latter fact
follows from computations in the next section. On the other hand,
there is a formal algebraic proof based on the construction of an explicit basis
of $\mathbf{A}_0$ given below.

Among the nine polynomials $t_k^{(p)}$, there are six containing $l_{3,3}^{(0)}$,
$l_{3,2}^{(0)}$, $l_{3,3}^{(1)}$, $l_{1,2}^{(1)}$, $l_{2,1}^{(1)}$ and
$l_{3,1}^{(1)}$ linearly, thus they can be eliminated. The remaining three
polynomials $t_k^{(p)}$ are used to reduce the power of $l_{1,1}^{(1)}$ (of
degree $4$) to $1$ and the powers of $l_{2,2}^{(1)}$ (of degree $4$) and
$l_{3,1}^{(0)}$ (of degree $3$) to 2. The remaining variables $l_{1,3}^{(1)}$
(of degree $2$), $l_{2,3}^{(1)}$ (of degree $3$) and $l_{3,2}^{(1)}$ (of degree
$5$) enter without restriction of power. This is in agreement with the
character~\eqref{chA0}. We shall call this basis the $\mathbf{A}_0$-basis. If we
need to compare two polynomials as elements of $\mathbf{A}_0$, we first bring
them to $\mathbf{A}_0$-basis subtracting elements of $\mathbf{F}\mathbf{A}$ and
compare them afterwards. We shall denote by  $\equiv$ equality in $\mathbf{A}_0$
(in other words, equality modulo $\mathbf{F}$).

\section{Separated variables}

Here we define the divisor in terms of $l(z)$ and vice versa. Divide $l(z)$ into
blocks
\begin{equation*}
  l(z)=\begin{pmatrix} a(z)&b(z)\\c(z)&d(z)\end{pmatrix}\,,
\end{equation*}
where $a(z)$ is of size $1\times 1$, which defines the shape of the other
entries.

Consider the eigenvector $\Psi(z,w)$, satisfying
\begin{equation*}
  (l(z)+wI)\Psi(z,w)=0\,.
\end{equation*}
The first component $\psi(z,w)$ of $\Psi(z,w)$  is a so-called Baker-Akhiezer
function. Notice that it is not a function on the spectral curve, but rather a
section of the linear bundle. The divisor
$\{(z_1,w_1),\ (z_2,w_2),\ (z_3,w_3)\}$ of zeros of $\psi(z,w)$ is well defined;
from a mechanical point of view it produces the separated variables. 
We impose the genericity assumption requiring that the points of the divisor are distinct.  Let us
express them in terms of the $l_{i,j}^{(p)}$.

For this introduce the polynomial
\begin{equation*}
  B(z)=\det\begin{pmatrix}b(z)\\ b(z)d(z)\end{pmatrix}=z^3+B_1z^2+B_2z+B_3\,.
\end{equation*}
It is easy to see that the zeros of this determinant coincide with projections
of the points of the divisor on the $z$-plane, i.e.\
\begin{equation*}
  B(z)=(z-z_1)(z-z_2)(z-z_3)\,.
\end{equation*}
These polynomials are in involution:
\begin{equation*}
  \{B(z),B(z')\}=0\,.
\end{equation*}
Explicitly, their coefficients are
\begin{align}\label{B}
  &B_1=2l^{(1)}_{1,2}+l^{(0)}_{3,3}l^{(1)}_{1,3}+l^{(1)}_{2,3}\,,\\
  &B_2=\(l^{(1)}_{1,2}\)^2+l^{(0)}_{3,3}l^{(1)}_{1,2}l^{(1)}_{1,3}
       -l^{(0)}_{3,2}\(l^{(1)}_{1,3}\)^2-l^{(1)}_{1,3}l^{(1)}_{2,2}
       +2 l^{(1)}_{1,2}l^{(1)}_{2,3}+l^{(1)}_{1,3}l^{(1)}_{3,3}\,,\nn\\
  &B_3=l^{(1)}_{1,2}l^{(1)}_{1,3}l^{(1)}_{3,3}-l^{(1)}_{1,2}l^{(1)}_{1,3}l^{(1)}_{2,2}
       +\(l^{(1)}_{1,2}\)^2l^{(1)}_{2,3}-\(l^{(1)}_{1,3}\)^2l^{(1)}_{3,2}\,.\nn
\end{align}

If $(z_i,w_i)$ is a point of the above divisor, the $3\times 2$ matrix
\begin{equation*}
  X(z_i,w_i)=\begin{pmatrix}b(z_i)\\d(z_i)+w_iI\end{pmatrix}
\end{equation*}
has rank one. Hence for arbitrary $2\times 3$ matrices $Y$ we find
\begin{equation*}
  \det (Y\cdot X(z_i,w_i))=0\,.
\end{equation*}
Setting
\begin{equation*}
  Y=\begin{pmatrix}1&0&0\\0&1&0\end{pmatrix},\quad
  Y=\begin{pmatrix}1&0&0\\0&-l^{(0)}_{3,2}&1\end{pmatrix},\quad
  Y=\begin{pmatrix}-l^{(0)}_{3,3}&1&0\\
                   -l^{(0)}_{3,2}-\(l^{(0)}_{3,3}\)^2&0&1\end{pmatrix},
\end{equation*}
one obtains respectively
\begin{align}
  &z_i^2+z_i\(l^{(1)}_{1,2}+l^{(1)}_{2,3}\)-w_il^{(1)}_{1,3}
   -l^{(1)}_{1,3}l^{(1)}_{2,2}+l^{(1)}_{1,2}l^{(1)}_{2,3}=0\,,\label{eq1}\\
  &w_iz_i+z_i\(l^{(1)}_{3,3}-l^{(0)}_{3,2}l^{(1)}_{1,3}-l^{(0)}_{3,3}l^{(1)}_{2,3}\)
   +w_i\(l^{(1)}_{1,2}+l^{(0)}_{3,3}l^{(1)}_{1,3}\)\,,\label{eq2}\\
  &+l^{(0)}_{3,3}l^{(1)}_{1,3}l^{(1)}_{2,2}-l^{(0)}_{3,3}l^{(1)}_{1,2}l^{(1)}_{2,3}
   -l^{(1)}_{1,3}l^{(1)}_{3,2}+l^{(1)}_{1,2}l^{(1)}_{3,3}=0\,,\nn\\
  &w_i^2+z_i\(l^{(0 )}_{3,2}l^{(1)}_{1,2}+l^{(0)}_{3,3}l^{(1)}_{2,2}
   +l^{(0)}_{3,2}l^{(0)}_{3,3}l^{(1)}_{1,3}+\(l^{(0)}_{3,3}\)^2l^{(1)}_{2,3}
   -l^{(1)}_{3,2}-l^{(0)}_{3,3}l^{(1)}_{3,3}\)\label{eq3}\\
  &+w_i\(l^{(1)}_{2,2}-l^{(0)}_{3,3}l^{(1)}_{1,2}-l^{(0)}_{3,2}l^{(1)}_{1,3}
   -\(l^{(0)}_{3,3}\)^2l^{(1)}_{1,3}+l^{(1)}_{3,3}\)
   +l^{(0)}_{3,2}\(l^{(1)}_{1,2}l^{(1)}_{2,3}-l^{(1)}_{1,3}l^{(1)}_{2,2}\)\nn\\
  &+\(l^{(0)}_{3,3}\)^2\(l^{(1)}_{1,2}l^{(1)}_{2,3}-l^{(1)}_{1,3}l^{(1)}_{2,2}\)
   +l^{(0)}_{3,3}\(l^{(1)}_{1,3}l^{(1)}_{3,2}-l^{(1)}_{1,2}l^{(1)}_{3,3}\)
   -l^{(1)}_{2,3}l^{(1)}_{3,2}+l^{(1)}_{2,2}l^{(1)}_{3,3}=0\nn\,.
\end{align}
These equations hold for $i=1,2,3$. They allow to express the polynomials in
$l_{i,j}^{(p)}$ on the left hand side as functions on $\Ja$, for example
\begin{equation*}
  l^{(1)}_{1,3}=\frac{\(1,z,z^2\)}{(1,z,w)}\,.
\end{equation*}

Let us show that they enable us to express all the $l_{i,j}^{(p)}$ as functions
on $\Ja$. For it we add one more equation:
\begin{equation}
  l^{(0)}_{3,2}+l^{(1)}_{1,3}+t_2^{(1)}=0\,.\label{eq4}
\end{equation}
From~\eqref{eq1} we find $l^{(1 )}_{1,2}+l^{(1 )}_{2,3}$ and $l^{(1 )}_{1,3}$,
and hence $l^{(0)}_{3,2}$ by~\eqref{eq4} in the form~\eqref{form}.
Equation~\eqref{eq2} gives us $l^{(1 )}_{1,2}+l^{(0 )}_{3,3}l^{(1)}_{1,3}$ in
the form~\eqref{form}. Since $l^{(0)}_{3,3}=t_1^{(1)}$ is a constant, we have
found $l^{(1)}_{1,2}$ and therefore $l^{(1)}_{2,3}$. Now by~\eqref{eq2}, the
element $l_{3,3}^{(1)}$ is expressed as quadratic in functions~\eqref{form}.
From~\eqref{eq3}, we obtain first $l_{2,2}^{(1)}$ and subsequently
$l_{3,2}^{(1)}$ again as quadratic in functions~\eqref{form}. Thus we have found
all the elements of the last two columns of $l(z)$ as functions on the affine
Jacobian. The elements of the first column are computed using
\begin{align*}
  &l^{(1)}_{1,1}+l^{(1)}_{2,2}+l^{(1)}_{3,3}-t_1^{(2)}=0\,,\\
  &l^{(0)}_{3,1}+l^{(1)}_{1,2}+l^{(1)}_{2,3}-t_3^{(1)}=0\,,\\
  &l^{(1)}_{2,1}-l^{(0)}_{3,3}l^{(1)}_{1,1}+l^{(0)}_{3,1}l^{(1)}_{1,3}
   -l^{(0)}_{3,3}l^{(1)}_{2,2}+l^{(0)}_{3,2}l^{(1)}_{2,3}+l^{(1)}_{3,2}
   +t^{(2)}_2=0\,,\\
  &l^{(1)}_{3,1}-l^{(0)}_{3,2}l^{(1)}_{1,1}+l^{(0)}_{3,1}l^{(1)}_{1,2}
   -l^{(0)}_{3,3}l^{(1)}_{2,1}-l^{(1)}_{1,3}l^{(1)}_{2,2}
   +l^{(0)}_{3,1}l^{(1)}_{2,3}+l^{(1)}_{1,2}l^{(1)}_{2,3}-t_3^{(2)}=0\,.
 \end{align*}

Motivated by~\eqref{eq1}, we introduce now the function
\begin{equation*}
  A(z)=\frac 1 {l^{(1)}_{1,3}}\(z^2+z\(l^{(1)}_{1,2}+l^{(1)}_{2,3}\)
       -l^{(1)}_{1,3}l^{(1)}_{2,2}+l^{(1)}_{1,2}l^{(1)}_{2,3}\)
\end{equation*}
such that
\begin{equation*}
  A(z_i)=w_i\,.
\end{equation*}
It can be shown that
\begin{align*}
  &\{A(z),A(z')\}=0\,,\\
  &\{A(z),B(z')\}=\frac 1 {z-z'}\big(zB(z')-z'B(z)\big)\,,
\end{align*}
which implies that the variables $\log{(z_i)}$, $w_i$ are canonically conjugated:
\begin{equation*}
  \{z_i,z_j\}=\{w_i,w_j\}=0,\qquad \{z_i,w_j\}=\delta_{i,j}z_i\,.
\end{equation*}
Notice that the coefficients of the function $A(z)$ are not holomorphic
functions on $\Ja$, but the symmetric functions of $w_i$ should be. Let us check
that.

The elementary symmetric functions of the variables $z_i$ are given by
\begin{equation*}
  \sigma _j(z_1,z_2,z_3)=(-1)^jB_j\,.
\end{equation*}
Next let us find the elementary symmetric functions of $w_j$.
Adding~\eqref{eq1} for $i=1,2,3$ we obtain
\begin{equation*}
  \sigma_1(w_1,w_2,w_3)
  =\frac 1 {l^{(1)}_{1,3}}\((B_1^2-2B_2)-B_1\(l^{(1)}_{1,2}+l^{(1)}_{2,3}\)
   -3l^{(1)}_{1,3}l^{(1)}_{2,2}+3l^{(1)}_{1,2}l^{(1)}_{2,3}\)\,.
\end{equation*}
Direct computation shows that the denominator $l^{(1)}_{1,3}$ cancels on the
right hand side. Then using~\eqref{eq3}, one finds $\sum w_i^2\,$;
also,~\eqref{eq2} allows to find $\sum z_iw_i$. Multiplying now~\eqref{eq3} by
$w_i$, summing over $i$ and using previous results, we find $\sum w_i^3$. So all
symmetric functions of $w_i$ are found, and they are indeed holomorphic
functions on the affine Jacobi variety.

\section{Cohomologies}

Only three $t^{(p)}_k$ do not lie in the centre of the Poisson algebra:
$t_2^{(2)}$, $t_3^{(2)}$, $t_3^{(3)}$. These are the integrals of motion.
They give rise to the three commuting vector fields
\begin{equation*}
  D_1x=\bigl\{t_2^{(2)},x\bigr\}\,,\quad D_2x=\bigl\{t_3^{(2)},x\bigr\}\,,
  \quad D_3x=\bigl\{t_3^{(3)},x\bigr\}\,.
\end{equation*}
It is easy to see that
\begin{equation*}
  D_i\mathbf{A}^{(p)}\subset \mathbf{A}^{(p+d(i))}\,,
\end{equation*}
where $d(i)$ is the degree of the $i$th vector field with
\begin{equation*}
  d(1) = 1\,,\quad d(2)=2\,,\quad d(3)=5\,.
\end{equation*}
Introduce the one-forms
\begin{equation*}
  d\sigma_i = \sum_{j=1}^3 d\sigma_i(P_j)
\end{equation*}
on $\Ja$ of degree $-d(i)$ and the differential $d=\sum D_id\sigma_i$. Define the
complex of $k$-chains
$\sum x_{i_1,\cdots i_k}d\sigma_{i_1}\wedge\cdots\wedge d\sigma_{i_k}$,
$x_{i_1,\cdots i_k}\in\mathbf{A}$ for $k=0,1,2,3$. We are interested in the
cohomologies of this complex with coefficients in the ring $\mathbf{F}$. This is
consistent, since
\begin{equation*}
  D_ix=0,\quad x\in \mathbf{F}\,.
\end{equation*}
It is easy to compute the $q$-Euler characteristic~\cite{NS,SZ,NS1}:
\begin{equation*}
\begin{split}
  \chi_q
  &=-\frac{\ch{(\mathbf{A}_0)}}{q^8(1-q)(1-q^2)(1-q^5)}\\
  &=-q^{-8}+q^{-7}-2q^{-5}+2q^{-3}-3q^{-2}-q^{-1}+2-q-3q^2+2q^3-2q^5+q^7-q^8\,.
\end{split}
\end{equation*}
This expression looks not very nice containing negative coefficients, but we
shall soon see that it has a rather natural interpretation. For the moment
notice that $\chi_1=-6$, which is the correct Euler characteristic for the
affine Jacobi variety. In the case under consideration, the theta divisor is
non-singular. Thus $\chi_1$ coincides with $(-1)^g g!$, which is the Euler
characteristic for an affine Abelian variety in generic position.

We have the one-forms
\begin{equation*}
\begin{split}
  d\tilde\sigma_i
   &=\sum_{j=1}^3 d\tilde\sigma_i(P_j)\\
  &=\frac{\big(\widetilde{Q}_i(z,w),z,w\big)}{(1,z,w)}d\sigma_1
    +\frac{\big(1,\widetilde{Q}_i(z,w),w\big)}{(1,z,w)}d\sigma_2
    +\frac{\big(1,z,\widetilde{Q}_i(z,w)\big)}{(1,z,w)}d\sigma_3
\end{split}
\end{equation*}
constructed via the second kind differentials, where
$\widetilde{Q}_1$, $\widetilde{Q}_2$, $\widetilde{Q}_3$ are given by~\eqref{Q}.
Working in $\mathbf{A}_0$, we can replace them by $5wz^2$, $2wz$, $z^2$
respectively. Using simple determinant identities, one shows that
\begin{align*}
  &d\tilde\sigma_i\wedge d\tilde\sigma_j
  =\frac{(\widetilde{Q}_i(z,w),\widetilde{Q}_j(z,w),w)}{(1,z,w)}\,
   d\sigma_1\wedge d\sigma_2\nn
   +\frac{(\widetilde{Q}_i(z,w),z,\widetilde{Q}_j(z,w))}{(1,z,w)}\,
   d\sigma_1\wedge d\sigma_3\\
  &\relphantom{d\tilde\sigma_i\wedge d\tilde\sigma_j=}
   +\frac{(1,\widetilde{Q}_i(z,w),\widetilde{Q}_j(z,w))}{(1,z,w)}
   \,d\sigma_2\wedge d\sigma_3\,,\\
  &d\tilde\sigma_1\wedge d\tilde\sigma_2\wedge d\tilde\sigma_3
  =\frac{(\widetilde{Q}_1(z,w),\widetilde{Q}_2(z,w),\widetilde{Q}_3(z,w))}{(1,z,w)}
   \,d\sigma_1\wedge d\sigma_2\wedge d\sigma_3\,.
\end{align*}

In order to express
\begin{equation*}
  \frac{\big(P_1(z,w),P_2(z,w),P_3(z,w)\big)}{(1,z,w)}
\end{equation*}
for any $P_j$ chosen from
$\{Q_1,Q_2,Q_3,\widetilde Q_1,\widetilde Q_2,\widetilde Q_3\}$, the simple
procedure below can be used. We start with the equations~\eqref{eq1},
~\eqref{eq2} and~\eqref{eq3}. One more equation can be obtained by multiplying
~\eqref{eq1} with $w_i$. Doing some linear combinations, we can derive from all
these equations the following ones:
\begin{align*}
  &L_{1,1}+z_iL_{1,2}+w_iL_{1,3}=\widetilde Q_1(z_i,w_i)\,,\\
  &L_{2,1}+z_iL_{2,2}+w_iL_{2,3}=\widetilde Q_2(z_i,w_i)\,,\\
  &L_{3,1}+z_iL_{3,2}+w_iL_{3,3}=\widetilde Q_{3}(z_i,w_i)\,.
\end{align*}
Here the $L_{i,j}$ are polynomials of $l^{(p)}_{k,l}$. Clearly all
$\frac{(P_1(z,w),P_2(z,w),P_3(z,w))}{(1,z,w)}$ are obtained as minors of the
matrix $||L_{k,l}||$.

Consider the space
\begin{equation*}
  V=\bigoplus_{i=1}^3\mathbb{C}d\sigma_i \ \oplus\
    \bigoplus_{i=1}^3\mathbb{C}d\tilde\sigma_i \,.
\end{equation*}
The spaces $\bigwedge^kV$ constitute cohomologies of the Jacobi variety. A part
of them is promoted to cohomologies of the affine Jacobi variety. Define
\begin{equation*}
  c=\sum d\sigma_i\wedge d\tilde\sigma_i\,.
\end{equation*}
Modular invariance of $C(P_1,P_2)$ implies that $c$ is the fundamental class of
the theta divisor. Hence
\begin{equation*}
  W^k=\(\wedge^kV\)/\(c\wedge\(\wedge^{k-2}V\)\)
\end{equation*}
are subspaces of the cohomology groups $H^k(\Ja)$.

Now we come to a tricky point. Since the theta divisor is non-singular, we might
use the theorem proven in~\cite{NS}, which states that the cohomology
groups $H^k$ coincide with $W^k$ for $k\le g-1$. But this theorem was shown for
cohomologies with coefficients in $\mathbb{C}$, while we are working with
coefficients in $\mathbf{F}$. There is no guarantee that the two things
coincide. But let us assume for the moment that this is indeed the case and see
what happens to $H^g$. Then we shall confirm our conclusions by direct
computation.

We find for the characters:
\begin{align*}
  &\ch{\(\wedge^0V\)}=1\,,\\
  &\ch{\(\wedge^1V\)}=q^{-5}+q^{-2}+q^{-1}+q+q^2+q^5\,,\\
  &\ch{\(\wedge^2V\)}=q^{-7}+q^{-6}+q^{-4}+2q^{-3}+q^{-1}+3+q+2q^3+q^4+q^6+q^7\,,\\
  &\ch{\(\wedge^3V\)}=q^{-8}+q^{-6}+2q^{-5}+q^{-4}+3q^{-2}+2q^{-1}+2q+3q^2+q^4
                      +2q^5+q^6+q^8\,.
\end{align*}
Obviously, $\ch{\(W^k\)}=\ch{\(\wedge^kV\)}-\ch{\(\wedge^{k-2}V\)}$. Now we
compute
\begin{equation*}
  \chi_q-\Big(1-\ch{\(W^1\)}+\ch{\(W^2\)}-\ch{\(W^3\)}\Big)=-1\,,
\end{equation*}
which implies two things: First, our assumption about the grading used seems to
be reasonable. Second, in addition to $W^3$ we seem to have only one element of
degree zero in the top cohomology. On the other hand, the top forms are
identified with $\mathbf{A}$ with the obvious prescription of degree
\begin{equation*}
  \deg{(x\,d\sigma_1\wedge d\sigma_2\wedge d\sigma_3)}=\deg{(x)}-8\,,
\end{equation*}
and the top cohomologies are identified with elements of $\mathbf{A}$ (or even
$\mathbf{A}_0$) up to derivatives. Then, if our description of cohomologies is
correct, we should have 15 such functions, and their degrees should correspond
to the character
\begin{equation*}
  q^8\(\ch{\(W^3\)}+1\)
  =1+q^2+q^3+q^4+2q^6+q^7+q^8+q^9+2q^{10}+q^{12}+q^{13}+q^{14}+q^{16}\,.
\end{equation*}

Suppose that the top cohomologies possess the basis
$h_{\al}d\sigma _1\wedge d\sigma _2 \wedge d\sigma _3$, where
$h_{\al}\in\mathbf{A}_0$. We identify with any $x\in\mathbf{A}_0$ an element of
$\mathbf{A}$ using $\mathbf{A}_0$-basis. Grading implies~\cite{NS}  that
\begin{equation*}
  x=\sum\limits_{\al}P_{\al}(D_1,D_2,D_3)h_\al
\end{equation*}
for any $x\in\mathbf{A}$, where $P_{\al}(D_1,D_2,D_3)$ has coefficients in
$\mathbf{F}$. This gives us a simple inductive procedure for computing the
top cohomologies. Provided the cohomologies up to degree $k-1$ have been found,
we act on all of them by all possible products of $D_j$ giving the total degree
$k$. We collect the results and develop them in $\mathbf{A}_0$-basis, obtaining
a matrix $M_k$. The dimension of the cohomologies of degree $k$ equals the
difference between the number of degree-$k$ elements of the $\mathbf{A}_0$-basis
and the rank of $M_k$. Then we choose as a representative of the cohomologies
the linearly independent elements of $\mathbf{A}_0$ which do not belong to the
image of $M_k$.

Proceeding this way we observe that the character of the top cohomologies is
$\ch{\(W^3\)}+1$, in agreement with the discussion above.

The following expressions can be taken to represent $h_{\al}$ except at degree
$8$. We set $\al=j,k$, with $j$ being the degree and $k$ counting elements of
the same degree. If there is only one element of a given degree, we omit the
second part. They are:
\begin{align}\label{simplecoh}
h_0&=1\,,&h_2&=\frac{(1,z,z^2)}{(1,z,w)}\,,&h_3&=\frac{(1,w,z^2)}{(1,z,w)}\\
h_4&=\frac{(1,w,zw)}{(1,z,w)}\,,&h_{6,1}&=\frac{(z,w,z^2)}{(1,z,w)}\,,
&h_{6,2}&=\frac{(1,z^2,zw)}{(1,z,w)}\,,\nn\\
h_7&=\frac{(z,w,zw)}{(1,z,w)}\,,&h_9&=\frac{(z,z^2,zw)}{(1,z,w)}\,,
&h_{10,1}&=\frac{(z,z^2,zw)}{(1,z,w)}\,,\nn\\
h_{10,2}&=\frac{(w,z^2,z^2w)}{(1,z,w)}\,,&h_{12}&=\frac{(z,z^2,z^2w)}{(1,z,w)}\,,
&h_{13}&=\frac{(w,z^2,z^2w)}{(1,z,w)}\,,\nn\\
h_{14}&=\frac{(w,zw,z^2w)}{(1,z,w)}\,,&h_{16}&=\frac{(z^2,zw,z^2w)}{(1,z,w)}\,.&\nn
\end{align}
Apparently, the corresponding cohomologies span $W^3$ as expected. Furthermore,
the functions~\eqref{simplecoh} have the simple denominator $(1,z,w)$, which
makes their expectation values factorisable.

The degree-zero subspace of $W^3$ is empty, but we computed that there is a
cohomology. According to the Remark from Section~\ref{reduction}, one can hope
that the corresponding function $h$ still has a simple pole $(1,z,w)$. We try to
take for $h$ one of the following functions:
\begin{equation*}
  \sigma_1(z)^2\frac{(1,z,z^2)}{(1,z,w)}\,,\quad
  \sigma_2(z)\frac{(1,z,z^2)}{(1,z,w)}\,,\quad \sigma_1(w)^2\,,\quad
  \sigma_2(w)\,.
\end{equation*}
However, it is easy to compute that modulo $\mathbf{F}$, these functions are 
linear combinations of five others, to wit
\begin{equation*}
  \frac{(1,z,z^4)}{(1,z,w)}\,,\quad\frac{(1,z^2,z^3)}{(1,z,w)}\,,\quad
  \frac{(1,zw,w^2)}{(1,z,w)}\,,\quad\frac{(1,zw^2,w)}{(1,z,w)}\,,\quad
  \frac{(z,w,w^2)}{(1,z,w)}\,.
\end{equation*}
The latter functions contain in their numerators one of the monomials $z^3$,
$z^4$, $w^2$, $zw^2$, which produce exact one-forms modulo $\mathbf{F}$. Hence,
for the representative of the cohomologies on level 8 we are forced to take an
expression with at least $(1,z,w)^2$ in the denominator.

It can be checked that the function
\begin{equation}
  h_8=\(\frac{(1,z,w^2)}{(1,z,w)}\)^2
  \label{h8}
\end{equation}
will do. This function has a clear denominator, but it is a little bit complicated in
terms of the $l_{i,j}^{(p)}$. On the other hand, it is easy to find out from
which degree in the $l_{i,j}^{(p)}$ the problem will start. Terms linear in
$l_{i,j}^{(p)}$ have degrees less than 8. For quadratic ones we observe that
neither of the four expressions
\begin{equation*}
  l^{(0)}_{3,1} l^{(1)}_{3,2}\,,\quad l^{(1)}_{2,3} l^{(1)}_{3,2}\,,\quad
  \(l^{(1)}_{2,2}\)^2,\quad l^{(1)}_{1,1}l^{(1)}_{2,2}
\end{equation*}
is cohomologically trivial, so neither of them can be reduced to a simple
denominator by adding total derivatives.

\vskip .2cm

\emph{Acknowledgements.}\; DM was supported by the LabEx ENS-ICFP:
ANR-10-LABX-0010/ANR-10-IDEX-0001-02 PSL grant. FS is grateful to Galileo Galilei Institute where this work was finished for
hospitality.

\end{document}